\documentclass[twocolumn,twoside]{revtex4}
\usepackage{graphicx,amsmath,multirow}
\usepackage{fancyhdr}
\usepackage[caption=false]{subfig}
\def\beq{\begin{equation}}
\def\eeq{\end{equation}}
\def\bea{\begin{eqnarray}}
\def\eea{\end{eqnarray}}
\def\nn{\nonumber}

\def \cL{{\cal L}}

\def \cM{{\cal M}}
\def \cA{{\cal A}}
\def \eps{\varepsilon}

\begin{document}

\title{FCNC in Concurrent Dark Photon and Dark $Z$ Models}


\author{Lopamudra Mukherjee}
\affiliation{Department of Physics \& Astronomy, University of Mississippi, Oxford, MS, USA, 38655}

\begin{abstract}
In this work we fit the available binned data of the differential decay distribution of the exclusive $B \to K^{(*)} \ell^+ \ell^-$ and $B_s \to \phi \mu^+ \mu^-$ decays to the mass and mixing parameters of a light dark vector boson model. Due to an incorrect assessment of the dominant contribution of the dark $Z$ model to the FCNC B-meson decays, a previous work in literature reported that $\mathcal{O}(1)$ mixings were allowed by the data. In this talk we report the correct calculations and constraints on the mixing parameters as well as the mass of the dark vector boson using the marginalization technique. We also study other relevant bounds on the parameter space from low energy experiments such as the atomic parity violation, $K^+ \to \mu^+ + invisible$ decay, $B_s - \overline{B}_s$ mixing etc and find that inspite of obtaining a good fit to the experimental $b \to s \ell^+ \ell^-$ data, the entire parameter space gets ruled out from some of the above bounds. In case of a model with tiny mixing and additional interaction of the dark $Z$ to the muon, some bounds are relaxed while some others are violated.
\end{abstract}

\maketitle

\thispagestyle{fancy}


\section{Introduction}
\begin{figure*}[t]
\centering
\includegraphics[width=135mm]{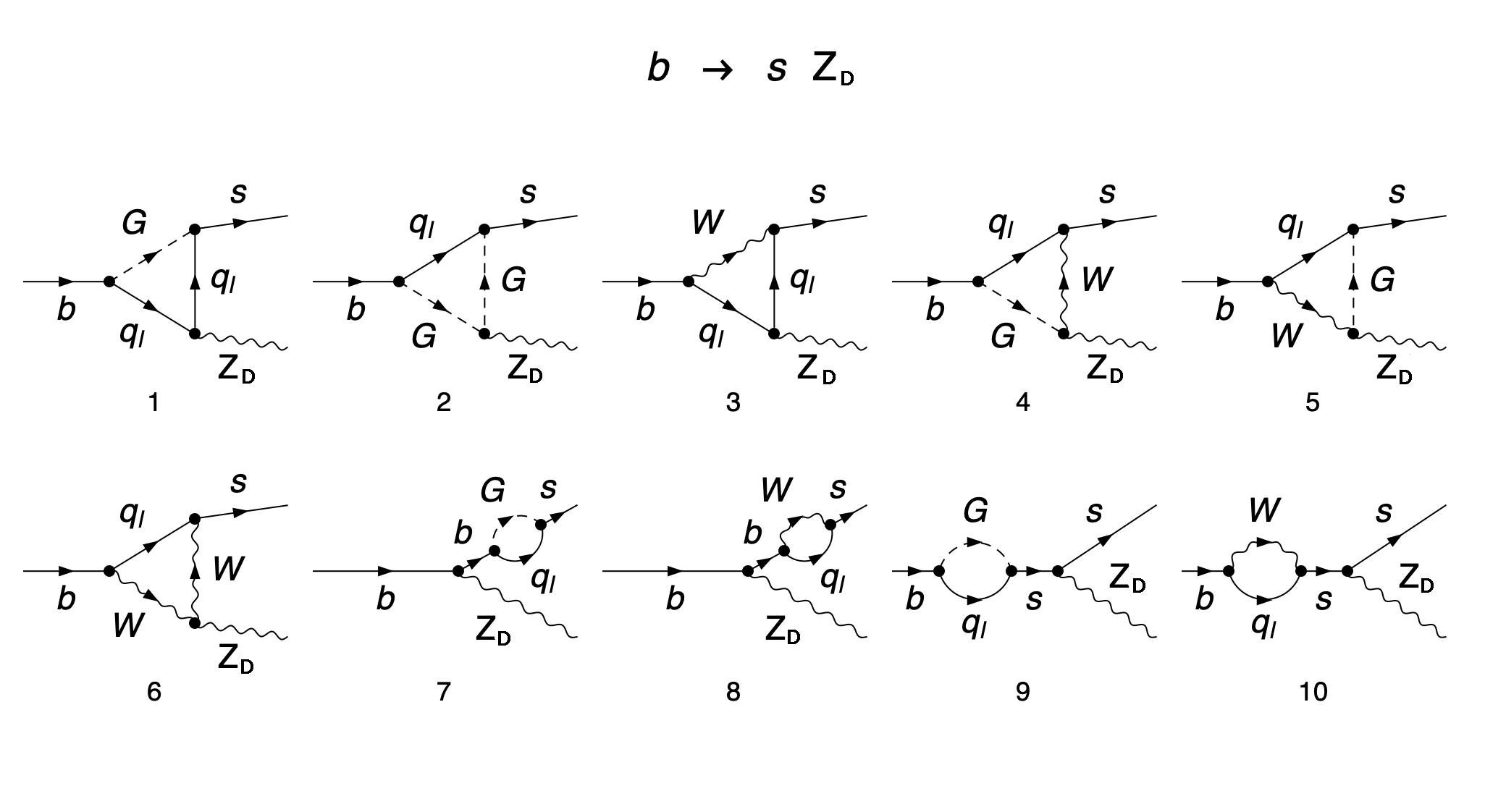}
\caption{Feynman diagrams for the FCNC process $b\to sZ_D$ at the parton level where the $G$'s are the Goldstone bosons.} \label{fig:feymnan-diag}
\end{figure*}

The anomalous results reported in the lepton flavour universality ratios $R_{K^{(*)}} = \mathcal{B}(B \to K^{(*)} \mu^+\mu^-)/\mathcal{B}(B \to K^{(*)} e^+e^-)$ have grabbed a lot of attention in the last decade. The extension of the SM by a dark gauged $U(1)$ symmetry with a dark TeV scale vector boson has been highly motivating in this regard with very few studies on light vectors \cite{Datta:2017pfz, Sala:2017ihs, Datta:2017ezo, Borah:2020swo,Crivellin:2022obd}. In this work \cite{ourpaper} we study a light vector boson belonging to a dark $U(1)_D$ gauge group in the mass range $0.01 < M_{Z_D}/\text{GeV} < 2$ which couples to the SM photon and $Z$ boson via kinetic and mass mixing thereby contributing to flavour changing neutral current (FCNC) decays. The gauge interaction Lagrangian in the model can be written as \cite{Davoudiasl:2012ag}
\beq
%\begin{split}
\cL_\text{gauge} =-\frac{1}{4} \left(B_{\mu\nu} B^{\mu\nu} + \frac{2\eps}{\cos\theta_W} B_{\mu\nu} Z_D^{\mu\nu} - Z_{D \mu\nu} Z_D^{\mu\nu}\right)
\label{eq:gauge-Lag}
\eeq
\normalsize
where $\eps$ is the kinetic mixing strength, $\theta_W$ is the Weinberg angle and
\beq
B_{\mu\nu} = \partial_\mu B_\nu - \partial_\nu B_\mu,Z_{D \mu\nu} = \partial_\mu {Z_D}_\nu - \partial_\nu {Z_D}_\mu.
%\end{split}
\eeq
After the diagonalization of the gauge sector, an induced coupling of the $Z_D$ to the SM electromagnetic current is generated, to leading order in $\eps,$
\beq
\mathcal{L}_D^{\text{em}} \supset e \varepsilon Z_D^\mu J_\mu^{\text{em}} -i e \varepsilon \left[\left[Z_D W^+ W^- \right]\right], 
\label{eq:lag1}
\eeq
which represents the ``dark photon" model. Further, the spontaneous breaking of the $U(1)_D$ symmetry makes $Z_D$ massive which can then mix with the SM $Z$ boson. The physical eigenstates can then be written in terms of the weak eigenstates as :
\bea
Z   &=  & Z^0 \cos\xi - Z_D^0 \sin\xi,  \nonumber\\
Z_D &=  & Z^0 \sin\xi + Z_D^0 \cos\xi, \
\label{eq:mass-mixing}
\eea
where
\bea
\tan 2\xi  &\simeq &  2\frac{M_{Z_d}}{m_Z} \delta = 2 \eps_Z \,
\label{eq:delta-epsZ}
\eea
parameterizes the mass mixing between the gauge bosons which then leads to
\beq
\mathcal{L}_D^{Z} \supset \frac{g}{\cos \theta_W} 
\varepsilon_Z Z_D^\mu J_\mu^{\text{Z}}-i g \cos \theta_W\varepsilon_Z \left[\left[Z_D W^+ W^- \right]\right]
\label{eq:lag2}
\eeq
in what is termed as the ``dark Z" model.
Eqs.~\eqref{eq:lag1} \& \eqref{eq:lag2} lead to new contributions to the $b \to s \ell^+ \ell^-$ via one loop penguin diagrams as shown in Fig.~\ref{fig:feymnan-diag}.

\section{Analysis}
\subsection{Calculation of the FCNC Amplitudes}
The dark photon and dark $Z$ contributions to the FCNC $B \to K^{(*)} \ell^+ \ell^-$ decay is similar to the SM photon and $Z$ mediated amplitudes with modified couplings. The hadronic part of the amplitude can be written as 
\bea
\footnotesize
& \cM_{Z_D}^{had} =  \langle K^{(*)} | \bar s \gamma_\mu P_{L/R} b | \bar B \rangle  \left[\left(\cA^{0,A}_{L/R}+\cA^{0,Z}_{L/R}\right) g^{\mu\nu}\right. \nn\\
&+ \left.\left(\cA^{2,A}_{L/R}+\cA^{2,Z}_{L/R}\right)\left(g^{\mu\nu}q^2-q^\mu q^\nu\right)\right]V_\nu^{Z_D} \\
&+   \langle K^{(*)} | \bar s iq_\mu \sigma^{\mu\nu} P_{L/R} b | \bar B \rangle  \left( M^{1,A}_{L/R}+M^{1,Z}_{L/R}\right) V_\nu^{Z_D} \nn
\label{DarkPhotonZ}
\eea 
in the effective Hamilton
\bea
H_{eff} &\ni& 
\left(\bar s \gamma^\mu P_{L/R} b \right)\left[g^{\mu\nu} \cA^{0}_{L/R} \right.
\nn \\  &+& \left.
\left(g^{\mu\nu}q^2-q^\mu q^\nu\right)\cA^{2}_{L/R}
\right]V_\nu \nn \\ &+& \left(\bar s \sigma^{\mu\nu}q_\mu P_{L/R}b \right) M^{1}_{L/R} V_\nu,
\eea
where  $P_{L/R}=(1/2)(1\mp\gamma^5)$, $g^{\mu\nu}$ is the metric tensor, $q_\mu$ is the outgoing momentum of a neutral vector boson $V$ and the hadronic current in terms of the form factors can be found in \cite{Bharucha:2015bzk}. In this model, $\cA^{0,A}_{L/R} = \cA^{0,Z}_R = \cA^{2,Z}_R = 0$ and the dominant contribution comes from the monopole term $\cA^{0,Z}_L$ of the dark $Z$ which was neglected in a previous calculation in literature \cite{Xu:2015wja}. Hence, we redo the analysis by correctly considering the dominant term.

\subsection{Calculation of the Wilson Coefficients}
In our model, we obtain new contributions to the Wilson coefficients (WCs) corresponding to the dimension six semileptonic operators $\mathcal{O}_9 = (\bar{s}_L \gamma^\mu b_L)(\bar{\ell} \gamma_\mu \ell)$ and $\mathcal{O}_{10} = (\bar{s}_L \gamma^\mu b_L)(\bar{\ell} \gamma_\mu \gamma_5 \ell)$ in terms of the hadronic loop factors for $\ell = e, \mu$ as : 
\footnotesize
\bea
\mathcal{C}_{9,\ell} &=& \left[\left(\cA^{0,Z}_{L} + \lbrace \cA^{2,A}_{L}+\cA^{2,Z}_{L}\rbrace q^2 \right) \times \right. \nn \\
&&  \left. \left(\frac{1}{q^2 - M_{Z_D}^2 + i \Gamma_{Z_D} M_{Z_D}}\right) \left(e\eps + \frac{g}{c_W}\eps_Z g_V^\ell \right) \right], \\
\mathcal{C}_{10,\ell} &=& \left[\left(\cA^{0,Z}_{L} + \lbrace \cA^{2,A}_{L}+\cA^{2,Z}_{L}\rbrace q^2 \right) \times \right. \nonumber\\
&&  \left. \left(\frac{1}{q^2 - M_{Z_D}^2 + i \Gamma_{Z_D} M_{Z_D}}\right) \left(\frac{g}{c_W}\eps_Z g_A^\ell \right) \right],
\label{eq:WC-NP}
\eea
\normalsize
where we neglect the dipole contributions which are tiny compared to the monopole ones and keep the $q^2$ and $\Gamma_{Z_D}$ dependences. Here $g$ is the SM weak coupling constant, $g_V^\ell = (-1 + 4 s_W^2)/2$ and $g_A^\ell = -1/2$.

\subsection{Dark Boson Decay Width}
\begin{figure}[h]
\centering
\includegraphics[width=80mm]{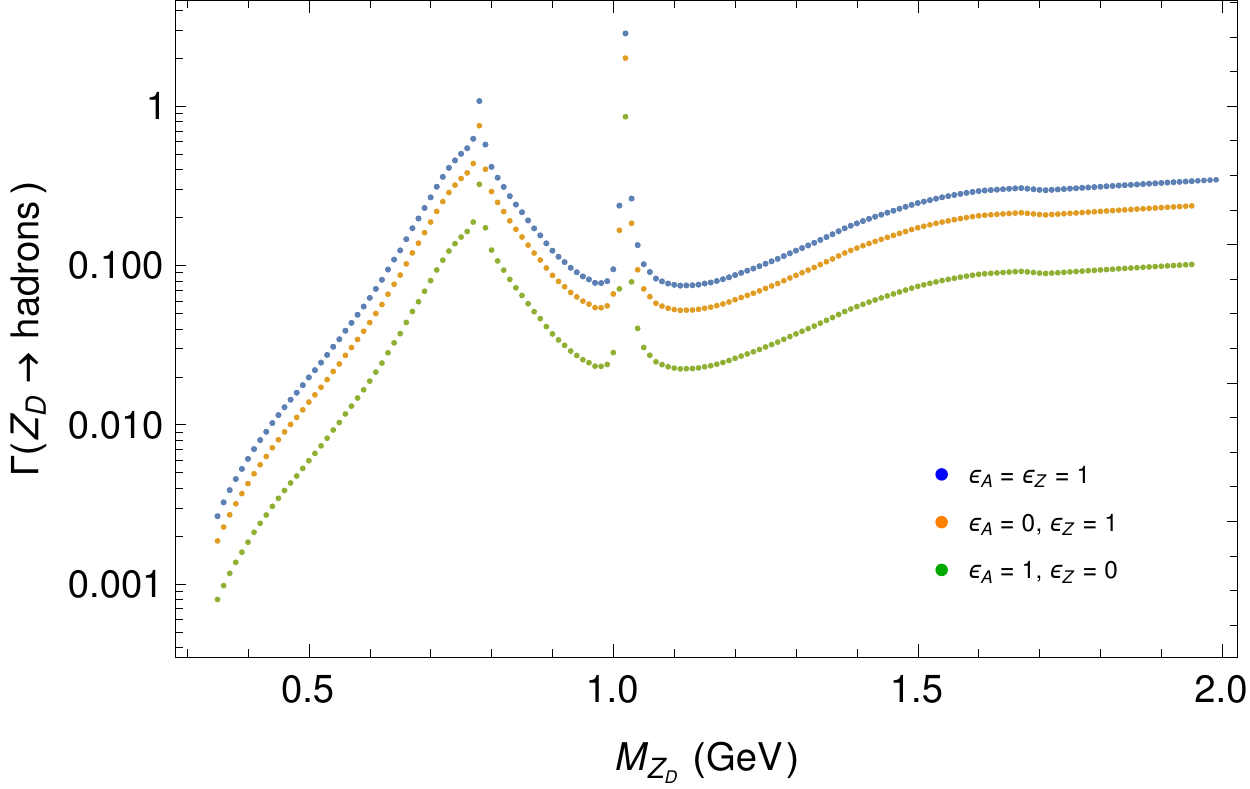}
\caption{Variation of the hadronic decay width of the dark bosons to light hadronic final states for different values of mixing parameters.} \label{fig:ZDhadron}
\end{figure}

The dark photon/$Z$ primarily decays to a pair of $e^+e^-$ and neutrinos, and to $\mu^+ \mu^-$ if kinematically allowed. For the analysis we allow both onshell and offshell decays of the dark boson to muons. Further, it can also decay to a pair of light hadrons such as $\pi^+ \pi^-, \pi^0 \pi^0$ etc when $500 \text{ MeV} \lesssim M_{Z_D} \lesssim 2$ GeV. But the hadronic decay width cannot be calculated from perturbative QCD approaches. We apply the data driven method used by \cite{Foguel:2022ppx} in our case and extend their analysis to calculate the hadronic decay width of the dark $Z$ which was not done before. We utilize the experimental measurement of the ratio of cross-sections $\mathcal{R}^\mathcal{H}_\mu = \sigma(e^+ e^- \to \mathcal{H})/\sigma(e^+e^- \to \mu^+ \mu^-)$ \cite{ParticleDataGroup:2020ssz} in order to approximate the hadronic decay width as
\beq
\Gamma(Z_D \to \mathcal{H}) = \Gamma(Z_D \to \mu^+ \mu^-) \times \mathcal{R}^\mathcal{H}_\mu .
\eeq 
In Fig.~\ref{fig:ZDhadron} we plot the hadronic decay width of the dark bosons for some benchmark values of the mixing parameters $\eps, \eps_Z$. 

\section{Data, Fit \& Constraints}
Our model has three free parameters : two mixing parameters ($\eps$ \& $\eps_Z$) and the mass $M_{Z_D}$. We fit them to the most recent experimental data on $B \to K^{(*)} \ell \ell$ and $B_s^0 \to \phi \mu^+ \mu^-$ transitions in different bins of $q^2$ below the $c\bar{c}$ resonance as listed in Table.~\ref{tab:data}. The SM and NP predictions for the differential branching fractions are calculated using \texttt{flavio}~\cite{Straub:2018kue}. 

\begin{table}[htp!!]
\begin{center}
\renewcommand{\arraystretch}{1.2}
\resizebox{\columnwidth}{!}{\begin{tabular}{|c|c|c|c|c|}
\hline
\textbf{Decay} & $\mathbf{q^2}$ \textbf{bin} &  \textbf{Experiment} & \textbf{SM Prediction} & $\mathbf{\chi^2_{SM}}$\\
\hline
\multirow{4}{*}{$\frac{d\mathcal{B}}{dq^2}(B^0 \to K^{*0} \mu \mu) \times 10^{8}$} & 0.1-0.98 & $11.06^{+0.67}_{-0.73}\pm 0.29 \pm 0.69$ \cite{LHCb:2016ykl} & $10.60 \pm 1.54$ & \multirow{23}{*}{76.92} \\
& 1.1-2.5 & $3.26^{+0.32}_{-0.31}\pm 0.10 \pm 0.22$ \cite{LHCb:2016ykl} & $4.66\pm 0.74$ & \\
& 2.5-4.0 & $3.34^{+0.31}_{-0.33}\pm 0.09 \pm 0.23$ \cite{LHCb:2016ykl} & $4.49\pm 0.70$ & \\
& 4.0-6.0 & $3.54^{+0.27}_{-0.26}\pm 0.09 \pm 0.24$ \cite{LHCb:2016ykl} & $5.02\pm 0.75$ & \\
\cline{1-4}
\multirow{3}{*}{$\frac{d\mathcal{B}}{dq^2}(B^+ \to K^{*+} \mu \mu) \times 10^{8}$} & 0.1-2.0 & $5.92^{+1.44}_{-1.30}\pm 0.40$ \cite{LHCb:2014cxe} & $7.97\pm 1.15$ & \\
& 2.0-4.0 & $5.59^{+1.59}_{-1.44}\pm 0.38$ \cite{LHCb:2014cxe} & $4.87\pm 0.76$ & \\
& 4.0-6.0 & $2.49^{+1.10}_{-0.96}\pm 0.17$ \cite{LHCb:2014cxe} & $5.43\pm 0.74$ & \\
\cline{1-4}
\multirow{6}{*}{$\frac{d\mathcal{B}}{dq^2}(B^+ \to K^{+} \mu \mu) \times 10^{8}$} & 0.1-0.98 & $3.32\pm 0.18 \pm 0.17$ \cite{LHCb:2014cxe} & $3.53 \pm 0.64 $ & \\
& 1.1-2.0 & $2.33\pm 0.15 \pm 0.12$ \cite{LHCb:2014cxe} & $3.53 \pm 0.58$ & \\
& 2.0-3.0 & $2.82\pm 0.16 \pm 0.14$ \cite{LHCb:2014cxe} & $3.51 \pm 0.52$ & \\
& 3.0-4.0 & $2.54\pm 0.15 \pm 0.13$ \cite{LHCb:2014cxe} & $3.50 \pm 0.63$ & \\
& 4.0-5.0 & $2.21\pm 0.14 \pm 0.11$ \cite{LHCb:2014cxe} & $3.47 \pm 0.60$ & \\
& 5.0-6.0 & $2.31\pm 0.14 \pm 0.12$ \cite{LHCb:2014cxe} & $3.45 \pm 0.53$ & \\
\cline{1-4}
\multirow{3}{*}{$\frac{d\mathcal{B}}{dq^2}(B^0 \to K^{0} \mu \mu) \times 10^{8}$} & 0.1-2.0 & $1.22^{+0.59}_{-0.52} \pm 0.06$ \cite{LHCb:2014cxe}& $3.28072 \pm 0.52272$ & \\
& 2.0-4.0 & $1.87^{+0.55}_{-0.49} \pm 0.09$ \cite{LHCb:2014cxe} & $3.25117 \pm 0.55614$ & \\
& 4.0-6.0 & $1.73^{+0.53}_{-0.48} \pm 0.09$ \cite{LHCb:2014cxe} & $3.20977 \pm 0.53957$ & \\
\cline{1-4}
\multirow{4}{*}{$\frac{d\mathcal{B}}{dq^2}(B_s^0 \to \phi \mu \mu) \times 10^{8}$} & 0.1-0.98 & $7.74 \pm 0.53 \pm 0.12 \pm 0.37$ \cite{LHCb:2021zwz} & $11.31 \pm 1.34$ & \\
& 1.1-2.5 & $3.15\pm 0.29 \pm 0.07 \pm 0.15$ \cite{LHCb:2021zwz} & $5.44 \pm 0.61$ & \\
& 2.5-4.0 & $2.34 \pm 0.26 \pm 0.05 \pm 0.11$ \cite{LHCb:2021zwz} & $5.14 \pm 0.73$ & \\
& 4.0-6.0 & $3.11 \pm 0.24\pm 0.06 \pm 0.15$ \cite{LHCb:2021zwz} & $5.50 \pm 0.69$ & \\
\cline{1-4}
\multirow{3}{*}{$\mathcal{B}(B^+ \to K^{+} ee) \times 10^{8}$} & 0.1-4.0 & $18.0^{+3.3}_{-3.0} \pm 0.5$ \cite{BELLE:2019xld} & $13.73 \pm 1.88$ & \\
& 4.0-8.12 & $9.6^{+2.4}_{-2.2} \pm 0.3$ \cite{BELLE:2019xld} & $14.11 \pm 1.88$ & \\
& 1.0-6.0 & $16.6^{+3.2}_{-2.9} \pm 0.4$ \cite{BELLE:2019xld} & $17.45 \pm 3.03$ & \\
\hline
\end{tabular}}
\caption{The experimental and SM predictions of the fit observables in bins of $q^2$ (GeV$^2$). The experimental measurement is taken from the reference listed in the third column while the SM predictions have been found out using \texttt{flavio}. The last column denotes the $\chi^2$ of all data in SM.}
\end{center}
\label{tab:data}
\end{table}

\begin{figure}[h]
\centering
\subfloat[]
{\includegraphics[scale=0.77]{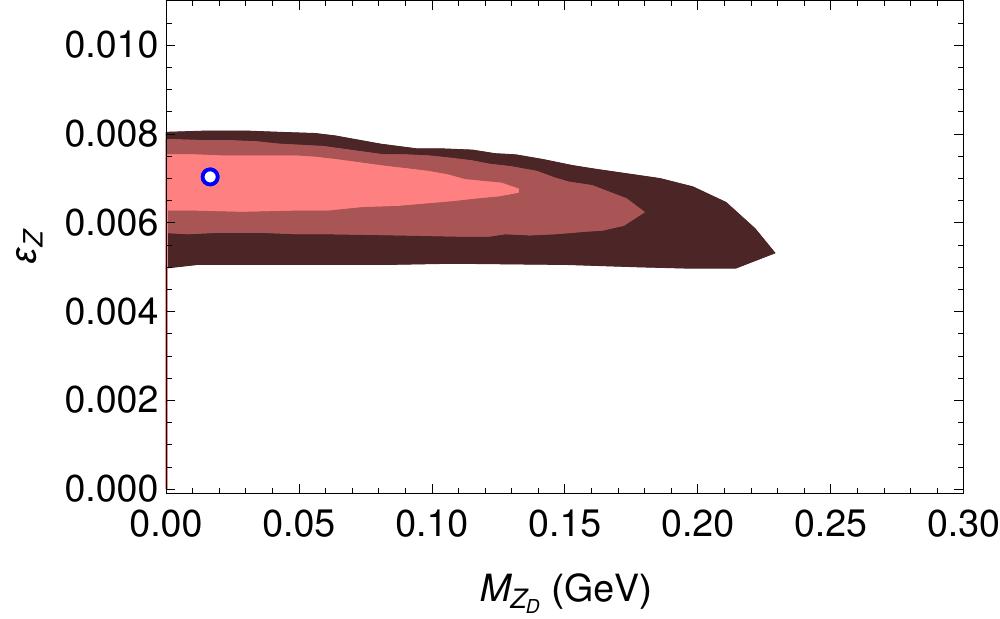}}\\
\subfloat[]
{\includegraphics[scale=0.745]{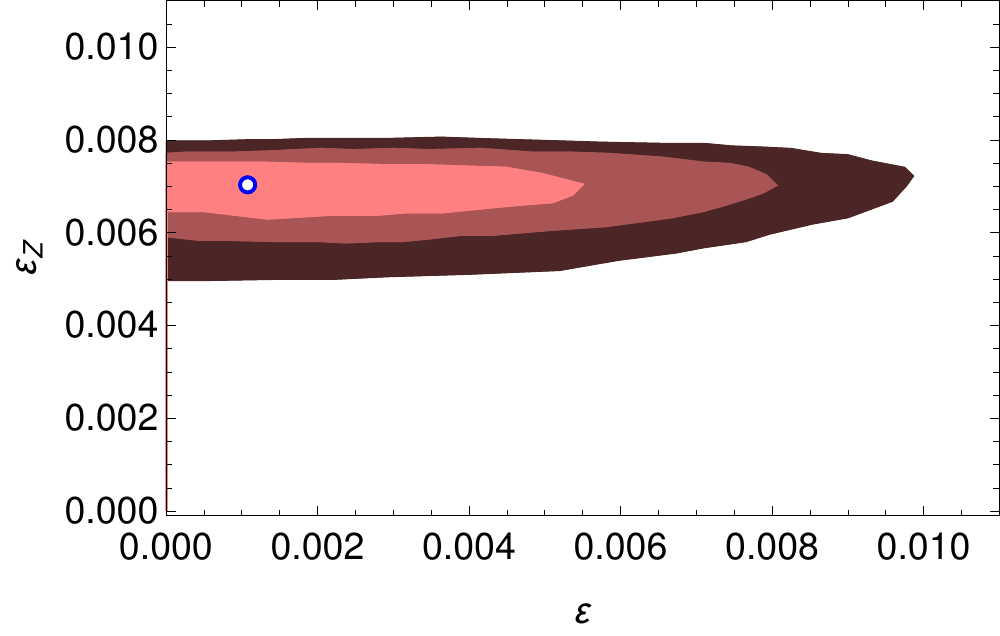}}
    \caption{The region in pink, brown and dark brown denote the constrained $1,2$ and $3 \sigma$ regions of the parameter space from the fit to the data in Table.~\ref{tab:data} with the best fit point depicted by the blue open circle.}
 	\label{fig:CaseA}
\end{figure}

For the dark vector boson model, we obtain a better solution with respect to the SM with $\chi^2_{NP} = 36.75$ at the best fit point where :
\beq
M_{Z_D} = 16.65~\text{MeV},~\eps = 0.00108,~ \eps_Z = 0.00703.
\label{eq:CaseA-bf}
\eeq 
We define $pull = \sqrt{\chi^2_{SM}-\chi^2_{NP}}$ which turns out to be $6.34$ for this case. Hence, we obtain a great improvement to the fit with respect to the SM for mixing parameters $\gtrsim \mathcal{O}(10^{-3})$ unlike the $\mathcal{O}(1)$ couplings obtained by authors \cite{Xu:2015wja}. We generate points on a grid and use the marginalization technique to obtain the constrained parameter space in $\eps - M_{Z_D}$ and $\eps_Z - \eps$ planes by marginalizing over the third parameter as shown in Fig.\ref{fig:CaseA}.

The parameter space depicted above is subject to strong constraints from several other low energy measurements which we discuss below :

\textbf{$\mathbf{B_s^0 - \overline{B}_s^0}$ Mixing :}
B-meson mixing plays an important role in the search for new physics. In this model, the dominant new physics (NP) correction to the mass difference comes from the monopole contribution of the dark $Z$ as given by
\footnotesize
\beq
 \Delta M_{B_s}^{NP} = \frac{1}{3}\frac{f_{B_s}^2 \hat B_{B_s} M_{B_s}}{M_{B_s}^2 - M_{Z_D}^2} \left(\cA^{0,Z}_L \right)^2 \left(1-\frac{5}{8}\frac{m_b^2}{M_{Z_D}^2}\right)
 \label{eq:delMs-NP}
\eeq
\normalsize
with an enhancement due to the longitudinal polarisation of $Z_D$. Due to this enhancement, the present uncertainty ($\sim 15\%$) in the theoretical estimate of $\Delta M_{B_s}$ restricts $\eps_Z \lesssim 0.001$ for $M_{Z_D} < 50$ MeV and could be at most $\sim 0.004$ for $M_{Z_D} \sim 250$ MeV. This rules out the entire parameter space obtained above.

\textbf{$\mathbf{B_s \to \mu^+ \mu^-}$ Decay :}
The rare decay $B_s \to \mu^+ \mu^-$ also puts tight constraints on the mixing parameter $\eps_Z$. The SM and experimentally measured rates are consistent with each other and can be found in \cite{Altmannshofer:2021qrr}. Since the NP rate depends only on the WC $C_{10,\mu}$, the constraint on $\eps$ is very weak. From our analysis, we find that the data prefers $\eps_Z < 0.005$ for $M_{Z_D}$ in the entire mass range of interest. Therefore, the constrained parameter space in Fig.~\ref{fig:CaseA} is once again ruled out.

\begin{figure}[h]
\centering
\includegraphics[width=80mm]{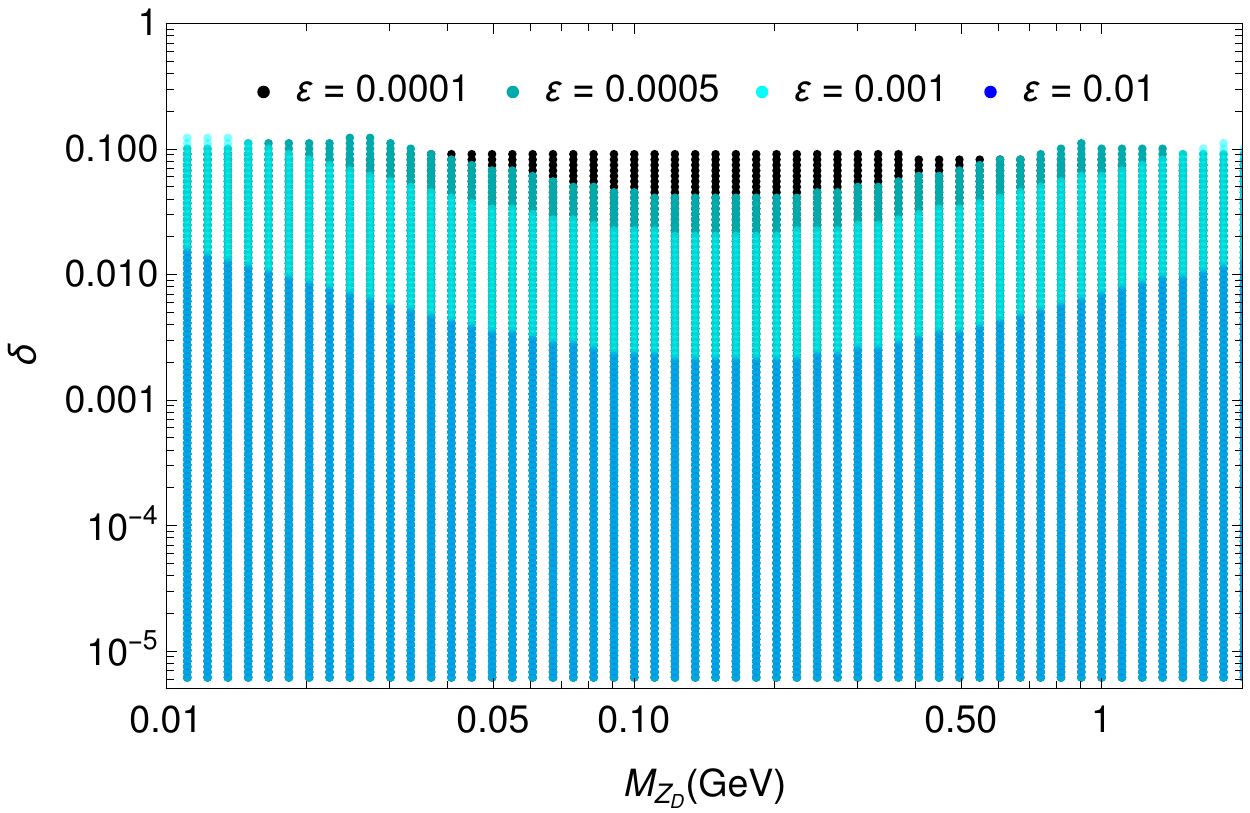}
\caption{The upper limit on $\delta$ as a function of the dark $Z$ mass from APV
observable measurements for different values of dark photon mixing parameter $\eps$.}
\label{fig:APV}
\end{figure}

\textbf{Atomic Parity Violation :}
Coupling of the dark bosons to the first generation fermions through mixing makes the model face stringent bounds from atomic parity violating (APV) observables such as the weak charges ($Q_W$) of proton and Caesium atom \cite{Cadeddu:2021dqx}. In Fig.~\ref{fig:APV} we show the allowed NP parameter space in the $\delta$-$M_{Z_D}$ plane for different values of $\eps$ when the uncertainty in the experimental data is taken within $3\sigma$ confidence limit (C.L.). Recall that $\delta$ is related to $\eps_Z$ as $\delta = m_Z \eps_Z/M_{Z_D}$. For $\eps \sim 10^{-4}$, $\delta \sim 0.1$ is favoured by the $3\sigma$ C.L. allowed data which can be translated to a bound on $\eps_Z$. We obtain a limit, $\eps_Z \lesssim 2\times 10^{-7}$ for $M_{Z_D} = 10$ MeV which makes the fit result in Eq.~\eqref{eq:CaseA-bf} in strong violation with this constraint.

\textbf{Other weak constraints :} The model parameters are also subject to other constraints such as the branching fractions of $B \to K^{(*)}\nu \bar{\nu}$ and $K^+ \to \pi^+ \nu \bar{\nu}$, the neutral Kaon mass difference due to mixing and the radiative $K^+ \to \mu^+ \nu_\mu Z_D $ which provide weaker bounds on the mixing parameters compared to the ones discussed above.

\section{Summary \& Outlook}
The dark vector boson model, despite providing a good solution to the $b \to s \ell \ell$ data, fails to abide by other important constraints. As an alternative, one can consider the case in which $Z_D$ has a direct vector-like coupling of strength $g_X$ to the muons and the mixing parameter $\eps_Z$ is very small but not zero such that it overcomes the strongest constraint coming from APV observables. We find that such a model provides a good fit to the $b \to s \ell \ell$ data but then fails to overcome constraints from the radiative $K^+ \to \mu^+ \nu_\mu Z_D$.

\begin{acknowledgments}
I would like to thank Alakabha Datta, Ahmed Hammad, Danny Marfatia and Ahmed Rashed for
collaboration on Ref.~\cite{ourpaper}. This work was funded by the National Science Foundation under Grant No. PHY-1915142.
\end{acknowledgments}

\bigskip % extra skip inserted

\end{document}